\documentclass[%
 amsmath,amssymb,
 aip,
 jcp,
 reprint,
 twocolumn,
]{revtex4-2}

\usepackage{graphicx}
\usepackage{dcolumn}
\usepackage{bm,color}
\usepackage{amsmath, amssymb, amsthm}
\usepackage{hyperref}

\begin{document}

\preprint{APS/123-QED}

\title{A~theory~for~diffusion-controlled~reactions~within~nonequilibrium~steady~states}

\author{Seokjin Moon}
\affiliation{Department of Chemistry, University of California, Berkeley, CA, 94720,  USA}
\author{David T. Limmer}
\email{dlimmer@berkeley.edu}
\affiliation{Department of Chemistry, University of California, Berkeley, CA, 94720, USA}
\affiliation{Kavli Energy NanoScience Institute, Berkeley, CA, 94720, USA}
\affiliation{\mbox{Materials Sciences Division, Lawrence Berkeley National Laboratory, Berkeley, CA, 94720, USA}}
\affiliation{\mbox{Chemical Sciences Division, Lawrence Berkeley National Laboratory, Berkeley, CA, 94720, USA}}

\date{\today}

\begin{abstract}
We study diffusion-controlled processes in nonequilibrium steady states, where standard rate theory assumptions break down. Using transition path theory, we generalize the relations between reactive probability fluxes and measures of the rate of the reaction. Stochastic thermodynamics analysis reveals how work constrains the enhancement of rates relative to their equilibrium values. An analytically solvable ion pairing model under a strong electric field illustrates and validates our approach and theory. These findings provide deeper insights into diffusion-controlled reaction dynamics beyond equilibrium.
\end{abstract}

\maketitle

\section{\label{sec:level1}Introduction}
Statistical rate theories connect the phenomenological laws of chemical kinetics of macroscopic systems to microscopic observables,\cite{peters2017reaction} leading to mechanistic understanding of reactions and enabling the development of efficient computational approaches to study them.\cite{singh2025variational,bolhuis2002transition,valsson2016enhancing,dickson2010enhanced} Here, we investigate diffusion-controlled processes in nonequilibrium steady states, where assumptions of most rate theories break down. Using transition path theory,\cite{vanden2006towards} we generalize the relationships between probability fluxes of reactive processes and physical observables, including mean first passage and transition path times. We provide thermodynamic interpretations using path reweighting in order to establish a relationship between the work done by the nonequilibrium force and the enhancement of reactive fluxes relative to those in equilibrium.\cite{kuznets2021dissipation} The theory we develop is validated by explicit studies of ion pairing under strong electric fields numerically and analytically.

Most rate theories rely on the assumption of timescale separation between slowly varying dynamical variables that relax a system globally and those faster variables that relax the system locally.\cite{limmer2024statistical, freidlin1998random} For example, Bennett--Chandler theory provides an efficient method for computing transition rates between two long-lived states and the rapid convergence of a dynamical correction to classical transition state theory is ensured by timescale separation.\cite{chandler1978statistical,frenkel2023understanding} However, this assumption breaks down in diffusion-controlled processes, where local relaxation is diffusive and inherently slow. Moreover, such diffusive dynamics can be easily driven far from equilibrium by external forces, in a manner not well described by perturbation theories around equilibrium.\cite{limmer2021large,gao2019nonlinear} Transition path theory\cite{vanden2006towards} (TPT) offers a  state-space-based framework for analyzing ensembles of reactive trajectories out of equilibrium. It applies generally to ergodic systems even when a phenomenological kinetic description of the dynamics may not be feasible,\cite{vanden2010transition} and can be generalized to periodically driven systems and non-stationary dynamics\cite{helfmann2020extending} as well as quantum mechanical systems.\cite{anderson2024coherent,anderson2023mechanism,brown2024unraveling}

Diffusion-controlled processes in nonequilibrium steady states occur across natural and synthetic systems. Recent advances in nanoscale fabrication techniques have enabled nanofluidic devices that confine an atomically thin channel of aqueous solutions between two-dimensional materials.\cite{bocquet2010nanofluidics,bocquet2020nanofluidics} In such strongly confined systems, the motion of electrolytes under an external electric field is distinct from bulk transport, reflecting stronger ion--ion correlations and enhanced ion pairing events.\cite{toquer2025ionic} The resulting nonlinear ionic current in response to a strong field is known as the Wien effect,\cite{onsager1934deviations,onsager1957wien} which results from modified fluctuations of the solution\cite{lesnicki2021molecular,lesnicki2020field} and changes of ion pairing rates\cite{onsager1934deviations, kaiser2013onsager} upon being driven far from equilibrium. Another important class of diffusion-controlled processes occurs in ligand--receptor interactions on cell surfaces. The association and dissociation kinetics of cells in hydrodynamic environments have been extensively studied, leading to the development of various models.\cite{chen2001selectin,dudko2008theory} For example, Bell's phenomenological law established an exponential dependence of the nonequilibrium dissociation rate to the shear stress,\cite{bell1978models} which can be understood within the context of recently reported dissipation bounds on the nonequilibrium rate enhancement.\cite{kuznets2021dissipation,kuznets2023inferring} However, most analysis of diffusive systems relies on simple chemical kinetic descriptions only valid for equilibrium, timescale separated systems. We go beyond these by developing a more general framework for analyzing such reactions.

The application and validation of generalized relations away from thermal equilibrium requires knowledge of key functions such as steady state distribution and committors. However, analytic expressions for these functions are often unknown in systems beyond one dimension.\cite{seifert2012stochastic} Here, we investigate an ion pairing system in the dilute limit. The Fokker--Planck equation and the backward Kolmogorov equation for this system have been extensively studied and their general solutions are known.\cite{isoda1994effect, hong1978solution, noolandi1979theory} We derive approximate extensions to these solutions that account for the finite size of ions by imposing judicious boundary conditions and compare them with results from simulations. Furthermore, we develop a method to compute the reactive flux and rates from simulations by defining specific correlation functions. The observed exponential behavior of the correlation functions allows us to apply path reweighting techniques,\cite{girsanov1960transforming,singh2025variational,keller2024dynamical} enabling the analysis of flux enhancement due to  nonequilibrium forces and their thermodynamic relationships. Our generalized relations extend the canonical theory of diffusion controlled reactions and illustrate how using traditional equilibrium relations\cite{hill2005free, lotz2018unbiased} within nonequilibrium steady states must be done with caution.

\begin{figure*}[ht]
\centering
\includegraphics[width=\textwidth]{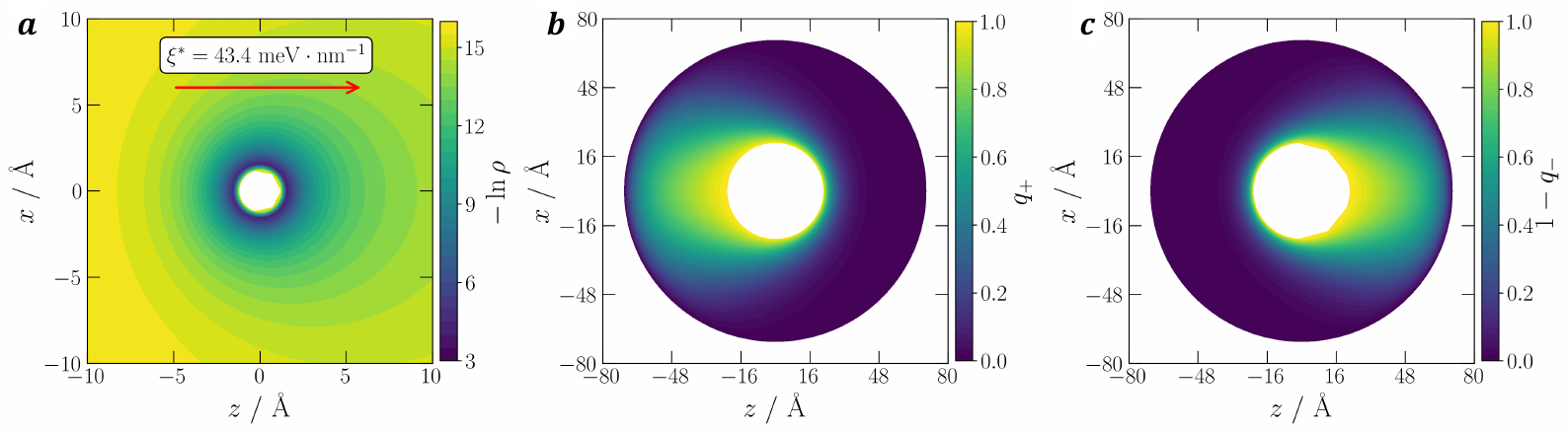}
\caption{\label{fig:1} a) A nonequilibrium free energy surface of an anion in cation-centered frame with the external field $\mathbf{\xi}^*= 43.4 \ \mathrm{meV} \cdot \mathrm{nm}^{-1}$ applied along the positive $z$ direction. b) The forward committor, $q_+$ and c) the complement of backward committor, $q_-$, for AB-reactive trajectories. The inner boundary location corresponds to the Bjerrum length $l_\mathrm{B}$ of the model system.}
\end{figure*}

\section{\label{sec:main1} Results and Discussions}
For concreteness, we consider the recombination of a cation and an anion in a structureless dielectric medium. Such an ion pairing process under strong electric fields in the dilute limit has been studied extensively since Onsager.\cite{onsager1934deviations} To simulate this process, we prepared a system composed of a cation--anion pair in a cubic box of length $L=200 \ \mathrm{\AA}$, interacting with each other via Coulomb and Lennard--Jones interactions. A constant external electric field was applied in the $z$ direction, leading to a nonconservative force $\mathbf{\xi}\hat{\mathbf{z}}$. Our study considered fixed values of $\xi$ in the range $0 \le \xi \le \xi^*$, with $\mathbf{\xi}^*= 43.4 \ \mathrm{meV} \cdot \mathrm{nm}^{-1}$ denoting the largest value considered. Periodic boundary conditions are applied in $x, y, z$ directions. At long times, the system approaches its nonequilibrium steady state due to the periodicity in the $z$ direction. Detailed model parameters are presented in Appendix \ref{AA}.

In the cation-centered frame, the motion of the anion can be modeled with an overdamped Langevin equation,\cite{turq1977brownian}
\begin{align}
    \dot{\mathbf{r}} = \beta D \left [ \mathbf{F}(\mathbf{r}) + \mathbf{\xi}\hat{\mathbf{z}} \right ] + \boldsymbol{\eta}_t \, ,\label{eq:langevin}
\end{align}
where $\mathbf{r}$ is the position of the anion relative to the cation, $\mathbf{F}$ is a conservative force from the pair potential, $\mathbf{\xi}\hat{\mathbf{z}}$ a non-conservative force of magnitude $\mathbf{\xi}$ and direction $\hat{\mathbf{z}}$, $\beta$ is the inverse of temperature times Boltzmann's constant, $D$ is the sum of diffusion coefficients of the cation and anion and $\boldsymbol{\eta}_t$ is Gaussian white noise of zero mean and $\langle \boldsymbol{\eta}_t \otimes \boldsymbol{\eta}_{t'}\rangle = 2D\mathbf{I}\delta (t-t')$, where $\mathbf{I}$ is the unit matrix.

The probability distribution function for the anion in steady state $\rho(\mathbf{r})$ satisfies the Smoluchowski equation,
\begin{equation}
    \begin{aligned}
    \mathcal{L}^{\dagger} \rho = -\beta D  \nabla & \cdot \left [ (\mathbf{F}+\mathbf{\xi}\hat{\mathbf{z}} )  \rho \right ] + D\nabla^2 \rho =0,\label{eq:sm}
    \end{aligned}
\end{equation}
where $\mathcal{L}^\dagger$ is the Fokker--Planck operator, and we assume period boundary conditions $\rho(\mathbf{r}+\mathbf{n}L ) = \rho(\mathbf{r})$ where $\mathbf{n}$ is an integer vector.
The general solutions of Eq.~\ref{eq:sm} are given in spherical coordinates,\cite{isoda1994effect}
\begin{equation}
    \begin{aligned}
        \rho(\mathbf{r})  = \left(\frac{l_\mathrm{B}}{2r}\right)^{\frac{1}{2}} &e^{-\frac{\beta}{2} V} \sum_{j=0}^{\infty } \left[\alpha_j Z_{1j}(r) + \beta_j Z_{2j}(r) \right] T_j (\mu) \, , \label{eq:gen}
    \end{aligned}
\end{equation}
where $r=|\mathbf{r}|$ is the distance, $\mu = \cos(\theta)$ is the polar angle from the positive $z$ axis, $\beta V = -l_\mathrm{B} / r - \beta \xi r\mu$ is the effective potential, 
$l_\mathrm{B} = \beta e^2/ 4\pi \varepsilon $ is the Bjerrum length,  $\varepsilon$ is the dielectric constant of the medium and $\alpha_j, \beta_j $ are linear coefficients to be determined by boundary conditions. Details on the two radial functions $Z_{1j}$, $Z_{2j}$ and angular functions $T_j$ can be found in Appendix \ref{AAA}.

Modifications to the analytic solution are required to get solutions corresponding to the molecular dynamics simulations. Firstly, we applied two radial boundary conditions,
\begin{equation}
    \begin{aligned}
        \rho(r_0, \mu) &= A \exp(\beta \mathbf{\xi} r_0 \mu ), \quad     \rho(R, \mu) = A'  \, , \label{eq:boundary}
    \end{aligned}
\end{equation}
where we set $r_0 = 1.2 \ \mathrm{\AA}$, $R = 173 \ \mathrm{\AA}$ and $A, \ A'$ are constants to be determined. The first boundary condition corresponds to the angular distribution of a contact ion pair in an external electric field, which is equivalent to an effective equilibrium system governed by a Boltzmann distribution. The second condition imposes a uniform angular distribution, valid if $R$ is much longer than any correlation length of the ions. The normalization condition of the probability distribution allows us to fix the coefficients if we have one more constraint. Here, we minimize the net steady state probability flux $\mathbf{J}_{ss} = \beta D(\mathbf{F}+\mathbf{\xi} \hat{\mathbf{z}})\rho  -D\nabla\rho$ over the dividing surface $S=\{\mathbf{r} \ : \ |\mathbf{r}| = 40 \ \mathrm{\AA} \}$ since it vanishes in the nonequilibrium steady state. Moreover, we modified the solution by applying radial masking functions in short and long radii regions to reproduce the radial histogram obtained from molecular dynamics simulations. The masking function incorporates the effect of the short-ranged Lennard--Jones interaction and the coordinate shift from spherical to cubic domain, the latter has minimal effects on the steady state distribution. The corresponding nonequilibrium free energy surface, defined as $ -\ln \rho(\mathbf{r})$, is shown in Fig.~\ref{fig:1}a. Under an applied field, the spherically symmetric free energy surface polarizes in the direction of the field, while the angular free energetics become uniform at long distances away from the cation at the origin.

In order to study recombination of the ions, we defined two subdomains with zero overlap, $A=\{\mathbf{r} : \: | \mathbf{r}| > 70 \ \mathrm{\AA} \}$ and $B = \{\mathbf{r} : | \mathbf{r}| < l_\mathrm{B}  \}$ corresponding to ion--separated and ion--paired states, respectively. Throughout, we use $l_\mathrm{B} = 22.3 \ \mathrm{\AA}$. Moreover, we defined $\Omega_{AB} = (A\cup B)^C$ to be the intermediate domain, and $\partial A$ and $\partial B$ as boundaries of the domains $A$ and $B$, respectively. 
In equilibrium, for rare events mediated by large barriers, the mechanism of a reaction is well described by the identification of a relevant transition state, or an ensemble of transition states.\cite{dellago2002transition} Away from equilibrium, or for diffusive dynamics, we can associate mechanistic insight with probabilistic generalizations defined within TPT. Specifically, the forward committor $q_+(\mathbf{r})$ is the probability of observing trajectories that reach $\partial B$ before touching $\partial A$ conditioned on starting at $\mathbf{r}$. The forward committor satisfies the backward Kolmogorov equation,
\begin{equation}
\begin{aligned}
    \mathcal{L} q_+ = \beta D(\mathbf{F}+\mathbf{\xi}\hat{\mathbf{z}}) \cdot \nabla &q_+ + D\nabla^2 q_+ = 0 \, ,
    \label{eq:forward}
\end{aligned}
\end{equation}
with boundary conditions $q_+ | _{A} = 0$ and $q_+ | _{B} = 1$, {\color{black}while} $\mathcal{L}$ is the adjoint of the Fokker--Planck operator. The forward committor is thus a measure of the progress of a reaction, analogous to a reaction coordinate. A complimentary quantity to the forward committor is the backward committor, $q_-(\mathbf{r})$, which is the probability of observing trajectories that came from $\partial A$ in the past, not from $\partial B$, conditioned on being at $\mathbf{r}$ at present. It satisfies the backward Kolmogorov equation under a time-dual dynamics, 
\begin{align}
    \tilde{\mathcal{L}} q_- = &-\beta D(\mathbf{F}+\mathbf{\xi}\hat{\mathbf{z}})\cdot \nabla q_- \\&+ 2D\nabla \ln \rho \cdot \nabla q_- + D \nabla^2 q_- = 0 \, ,\nonumber \label{eq:backward}
\end{align}
with boundary conditions $q_- | _{A} = 1, \quad q_- | _{B} = 0$. The generator of time-dual dynamics is denoted $\tilde{\mathcal{L}}$ and $\rho$ is the solution of the Fokker--Planck equation in steady state. The time-dual dynamics is a conjugate dynamics whose Fokker--Planck current has an inverted direction everywhere. In equilibrium systems where $\mathbf{\xi}=0$, $\nabla \ln \rho  = \beta \mathbf{F}$ and $\mathcal{L} = \tilde{\mathcal{L}}$. As a consequence in equilibrium, $q_+ = 1-q_-$.\cite{vanden2010transition}

The forward committor has a general solution for purely Coulomb-interacting systems,\cite{noolandi1979theory} and is of the same form as the stationary distribution in Eq.~\ref{eq:boundary} except for an additional Boltzmann factor, $q_+ = \exp(\beta V) \rho $, where the boundary conditions fully determine all linear coefficients. The solution is shown in Fig.~\ref{fig:1}b. While in equilibrium, ions are equally likely to react across all $\theta$ due to the orientational symmetry of the steady state, under an applied field, $q_+(\mathbf{r})$ is appreciable only for a narrow range of angles near $\theta \approx \pi$. The most likely configurations for a recombination event {\color{black}at large fields} are when the ions are displaced along the direction of the applied field in such a way that the field can convect the ions towards each other. 

The backward committor is related to the steady state generated under boundary conditions that ensure reactivity. We let $\rho_{AB}(\mathbf{r})$ be the steady state solution of the Smoluchowski equation,
\begin{equation}
    \begin{aligned}
    \mathcal{L}^{\dagger} \rho_{AB} = 0, \label{eq:kramers}
    \end{aligned}
\end{equation}
associated with a source $\rho_{AB}(\mathbf{r} ) |_{\partial A} = \rho(\mathbf{r} )$ that maintains a steady state distribution of reactant species, and a sink $\rho_{AB}(\mathbf{r}) |_{\partial B }= 0$ that removes products. It can be verified that $\rho_{AB} = \rho q_-$. With the general solutions in Eq.~\ref{eq:gen}, we computed the backward committor and the complement, $1-q_-$, is shown in Fig.~\ref{fig:1}c. This complement is clearly different from the forward committor, as a consequence of the breaking of time-reversal symmetry. It is appreciable around $\theta \approx 0$, indicating the most likely configurations for dissociation are at the opposite side of the recombination in state space. Away from equilibrium, the mechanisms of forward and backward transitions need not be the same.\cite{dickson2009separating,heller2024evaluation}

\subsection{Characteristic Frequencies of Recombination}
The committors and steady state probability distribution are closely related to the measure of how frequently the reactive processes occurs in steady state. We define reactive trajectories by the ordered family of times $\{(t_j^- , t_j^+ )\}$ with integer $j$ satisfying,
\begin{align}
    X_{t_j^-} \in \partial A, \quad X_{t_j^+} \in \partial B, \quad  \: X_t \in \Omega_{AB}\, , \label{eq:reactive}
\end{align}
where $\{ X_t \}_{t \in (t_j^-, t_j^+)}$ denotes the ensemble of short trajectories that fulfill this conditioning, and are referred to as the ensemble of AB-reactive trajectories. For a long trajectory of an ergodic system, we can define the mean frequency of observing AB-reactive trajectories, $\nu_{AB}$ by counting the number of $j$ satisfying Eq.~\ref{eq:reactive} divided {\color{black}} by the duration of trajectory. TPT offers a connection between the mean frequency $\nu_{AB}$ and the probability flux associated with the solution of Eq.~\ref{eq:kramers}, $\mathbf{J}_{AB, \xi} = \beta D(\mathbf{F} + \mathbf{\xi}\hat{\mathbf{z}})\rho_{AB} -D \nabla \rho_{AB}$,
\begin{equation}
    \begin{aligned}
    \nu_{AB}(\xi) = & \int_{S} d\sigma_S (\mathbf{r}) \hat{\mathbf{n}}_S(\mathbf{r}) \cdot \mathbf{J}_{AB,\xi}(\mathbf{r})   \, ,
    \end{aligned} \label{eq:flux}
\end{equation}
where $S \subset \Omega_{AB}$ is an arbitrary dividing surface, $\hat{\mathbf{n}}_S$ is the unit normal vector pointing to $B$, and $d\sigma_S$ is the surface element of $S$. We note that the mean frequency $\nu_{AB}$ can also be written in terms of committors and steady state probability.\cite{vanden2006towards}

From molecular simulations, the mean time to observe AB reactive trajectories can be measured by defining a time correlation function with specific conditioning. Here, we define the correlation function $h_\xi(\tau)$,
\begin{align}
    h_\xi(\tau) = \langle I\left( \{ X_t\}  \right)  \rangle_\xi\, , \label{eq:tcf}
\end{align}
where the average is taken over a trajectory ensemble of length $\tau$ under field $\xi$, $I$ is an indicator functional of trajectory which is 1 if the trajectory contains any AB reactive trajectories, and 0 otherwise. From $h_\xi(\tau)$, the mean time for observing an AB reactive trajectory, $\langle t_{AB}\rangle_\xi$ is readily obtained by
\begin{align}
    \langle t_{AB}\rangle_\xi = \int_{0}^{\infty} dt \: t \frac{dh_\xi(t)}{dt} = \int_0^\infty dt \: \left[ 1-h_\xi(t) \right]\, , \label{eq:meantime}
\end{align}
where we assume that $\lim_{t\rightarrow \infty} t(1-h_\xi(t)) =0$ which is valid if reactive trajectories are rare. 

\begin{figure}
\centering
\includegraphics[width=8.3cm]{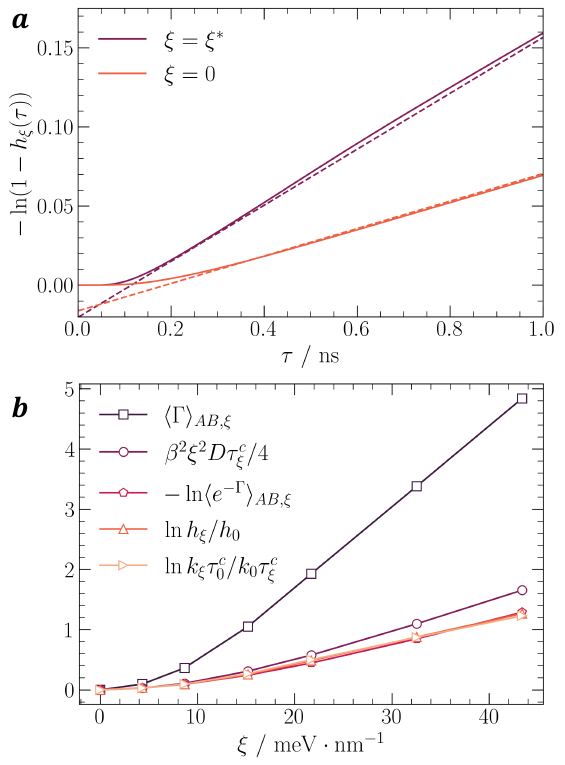}
\caption{\label{fig:2}a) Reactive flux correlation function with and without the external force $\mathbf{\xi}= \xi^*$ (solid lines) and a fit to a delayed exponential (dashed lines).  b) Estimates of the flux enhancement from trajectory reweighting, evaluated at observation time $\tau^*=2\tau^c_\xi$.}
\end{figure}

As shown in Fig.~\ref{fig:2}a, the correlation functions with and without the field change exponentially over intermediate times, a sign of a rare event, though the exponential behavior begins after a finite time lag consistent with prior work on diffusion controlled reactions.\cite{szabo1980first,dudko2005time} We model this behavior as a delayed exponential function,
\begin{equation}
h_\xi(\tau) =
\begin{cases} 
    1- e^{-k_\xi(\tau-\tau_\xi^{c})} , & \text{if } \tau \geq \tau^c_\xi  \\ 
    0, & \text{otherwise} \label{eq:dem}
\end{cases}
\end{equation}
where $k_\xi$ is the typical frequency of the rare event and $\tau^c_\xi$ is the mean transition path time, the averaged  time a reactive trajectory spends in $\Omega_{AB}$. Using this model, we compute,
\begin{align}
    \langle t_{AB}\rangle_\xi =  \int_{\tau_\xi^c}^\infty dt \: k_\xi t e^{-k_\xi(t-\tau_\xi^c)} = \tau^c_\xi + k_\xi^{-1}\, , \label{eq:meandem}
\end{align}
for which we find an increase in $k_\xi$ with increasing $\xi$, and a decrease in $\tau_\xi^c$ with increasing $\xi$. For $\xi^* = 43.4 \ \mathrm{meV \cdot nm^{-1}}$,  $\nu_{AB}^{-1}(\xi^*) = 5.68 \ \mathrm{ns}$, which agrees well with the characteristic time to react estimated from $\langle t_{AB} \rangle_{\xi^*} = 5.68 \ \mathrm{ns}$ and  $k_{\xi^*}^{-1} = 5.56 \ \mathrm{ns}$. Therefore, we confirm that the correlation function defined in Eq.~\ref{eq:tcf} behaves like a traditional side--side correlation function of typical barrier crossing system,\cite{chandler1978statistical} leading to a general kinetic framework for nonequilibrium diffusion-controlled processes.

The exponential behavior of the correlation function allows us to connect the ratio of correlation functions to Kramers' reactive flux, $\nu_{AB}(\xi)$. Using the model correlation function in Eq.~\ref{eq:dem}, the log ratio of the correlation functions at $\tau^* = 2\tau^c_{\xi}$, a sufficiently short time while ensuring exponential behavior, can be written as,
\begin{equation}
    \begin{aligned}
        \ln \frac{h_\xi (\tau^*)}{h_0 (\tau^*) } = \ln \frac{k_\xi}{k_0} + \ln \frac{\tau^c_{ 0}}{\tau^c_{ \xi}} +O\left( (1-\frac{\tau^c_{ 0}}{\tau^c_{ \xi}})^2\right) \, , \label{eq:cf2rates}
    \end{aligned}
\end{equation}
and both sides of the equality are shown in Fig.~\ref{fig:2}b. As a function of $\xi$, the measures of the frequency of recombination are found to increase. The ratio of mean transition path times contributes to the relation in diffusion-controlled process, while it becomes negligible in barrier crossing events where the transition paths are fleeting.\cite{kuznets2021dissipation}

\subsection{Thermodynamics of the Field-dependent Reactive Flux}

In order to interpret the field dependence of the flux enhancement thermodynamically, we follow recent work on trajectory reweighting,\cite{kuznets2021dissipation} generalized to a nonequilibrium diffusion-controlled process. In particular, the relative likelihood of observing a trajectory $ \{X_t\}$ of duration $\tau$ given an initial condition $\mathbf{r}(0)$ in the presence of different forces can be computed from a Girsanov transform.\cite{girsanov1960transforming} The log ratio of path probabilities with external field, $P_\xi [ \{X_t\} | \mathbf{r}_0 ]$, and without external field, $P_0 [ \{X_t\} | \mathbf{r}_0 ]$ is
\begin{equation}
\begin{aligned}
        &  \Gamma'(\{ X_t \} | \mathbf{r}_0) := \ln  \left( P_\xi [ \{X_t\} | \mathbf{r}_0 ]/P_0[\{X_t\} | \mathbf{r}_0 ] \right)  \\
        &= -\frac{\beta D}{4}  \int_{0}^{\tau} dt \left [ \mathbf{\xi}^2 - 2\mathbf{\xi} \hat{\mathbf{z}}\cdot ( \beta^{-1} D^{-1}\dot{\mathbf{r}} -  \mathbf{F}) \right ] \, , \label{eq:com}
\end{aligned}
\end{equation}
which follows from the difference of stochastic actions for the overdamped diffusion interpreted in the Ito sense.\cite{gao2019nonlinear} This expression allows us to write the ratio of correlation functions of two different trajectory ensembles into a conditioned trajectory ensemble average,
\begin{equation}
\begin{aligned}
   \ln \frac{h_\xi (\tau)}{h_0 (\tau) }  &= - \ln \frac{\left \langle e^{-\beta \Gamma'(\{X_t\} | \mathbf{r}_0 )} \frac{\rho_0(\mathbf{r}_0)}{\rho_\xi (\mathbf{r}_0)} I(\{X_t\}) \right  \rangle_\xi}{\langle I(\{X_t\})  \rangle_\xi}  \\
   &:= -\ln \left \langle e^{-\Gamma} \right \rangle_{AB, \xi} \, ,\label{eq:reweight}
\end{aligned}
\end{equation}
where we have introduced a functional $\Gamma (\{X_t\} | \mathbf{r}_0 ) := \beta \Gamma'(\{ X_t \} | \mathbf{r}_0) + \ln \rho_\xi (\mathbf{r}_0)/ \rho_0 (\mathbf{r}_0)$ and define a conditioned average in the $AB$ reactive ensemble, $\langle \dots \rangle_{AB,\xi} := \langle \dots I(\{X_t\})\rangle_\xi / \langle I(\{X_t\}) \rangle_\xi $. This relation is exact for any $\tau$ and $\xi$, though the exponential average at long $\tau$ requires exponentially many samples to converge. For this system, it is possible to converge the exponential from brute-force simulations, and this is shown in Fig.~\ref{fig:2}b, and in expected agreement with the ratio of correlation functions.

The convexity of the exponential function allows us to apply Jensen's inequality to Eq.~\ref{eq:reweight},
\begin{align}
    -\ln \langle e^{-\Gamma}  \rangle_{AB, \xi}  \le  \langle \Gamma  \rangle_{AB, \xi}\, , \label{eq:jensen} 
\end{align}
from which we can deduce a simple thermodynamic interpretation by decomposing the change in stochastic action.  The first term in Eq.~\ref{eq:com}, proportional to $\xi^2$, is interpreted as the work done by the field for a free ionic system $W_{f}(\tau) = \beta D \mathbf{\xi}^2 \tau$. The second term is the work due to the field defined from stochastic thermodynamics\cite{seifert2012stochastic} as $W(\{X_t \}) := \mathbf{\xi} \int dt \ \hat{\mathbf{z}} \cdot \dot{\mathbf{r}} $. The final term $\beta^2D \xi \int dt \  \hat{\mathbf{z}}\cdot \mathbf{F}/2 $ is identified as the excess dynamical activity.\cite{kuznets2021dissipation}  Within our definition of domain boundaries, where $\partial B$ is located in the vicinity of the Bjerrum length, this term is negligible, since the conservative forces are small. With these identifications, the ratio of reactive frequencies becomes
\begin{align}
    \ln \frac{k_\xi}{k_0} \frac{\tau^c_{ 0}}{\tau^c_{ \xi}}  \le \frac{\beta}{2}\langle W  \rangle_{AB, \xi} -  \frac{\beta}{4} W_f   + \left \langle \ln \frac{\rho_\xi }{\rho_0}  \right \rangle_{AB, \xi}\, , \label{eq:Jensen}
\end{align}
which sums the two work contributions with the ratio of initial conditions with and without the field. The inequality is shown in Fig.~\ref{fig:2}b for a range of applied fields, evaluated at $\tau^* = 2\tau^c_{ \xi}$. The inequality shows how the mean work done on the system conditioned on $AB$ reactive trajectories bounds the enhancement of the reactive flux. 

The deviation from the saturation of the bound can be attributed to the broadening of the conditioned work distribution, which physically corresponds to the field inefficiently promoting recombination. Linearization of the exponential average removes correlations between the initial probability ratio and the work distribution, which also contributes to the weakening of the bound. To separate contributions from broadening and decorrelation, we applied Jensen's inequality only to the work and ignored the initial ratio contribution, in Fig.~\ref{fig:s1}a in Appendix \ref{AB}. The broadening of the work distribution contributes to the weakened bound in all field cases, while including the initial density correlation tends to strengthen the bound. The initial density correlation is negligible in the small field regime, while it becomes significant for high fields.

In the limit of free ion recombination in an external field, we expect the conditioned average of the work to be equal to the free work. This is a consequence of the ions largely freely diffusing in the intermediate domain between reactants and products. Correspondingly, we find a reasonable approximation to the flux enhancement is given by 
\begin{equation}
\ln \frac{k_\xi}{k_0} \frac{\tau^c_{ 0}}{\tau^c_{ \xi}} \approx \beta^2 \xi^2 D \tau^c_\xi/4\, ,
\end{equation}
which means that the recombination flux enhancement is determined just by the work done by an external field on a free charge applied over a characteristic mean transition path time. This approximation works very well even for relatively high field cases, providing a useful model to estimate rate enhancement of a Coulomb interacting system under moderate applied electric fields.

\subsection{Mean First Passage and Transition Path Times}

Kramers' theory provides a way to compute rate constant as the probability flux over the population of reactants. In typical barrier crossing events, the inverse of the Kramers' rate constant equals the mean first passage time, $\tau_\xi(\mathbf{r})$, defined as the average time it takes for trajectories starting at $\mathbf{r}\in A$ to reach the boundary $\partial B$. Importantly, the mean first passage time is insensitive to the exact choice of $\mathbf{r}$ within $A$. We generalize Kramers' theory to extend its applicability to nonequilibrium steady states of diffusion-controlled processes. The mean first passage time satisfies the following differential equation,\cite{hanggi1990reaction}
\begin{align}
    \mathcal{L} \tau_\xi =  -1,   \label{eq:DEmfpt}
\end{align}
with boundary condition $\tau_\xi |_{\partial B} = 0$. We find that the mean first passage time is related to the reactive flux in a nonequilibrium steady state when evaluated on $\partial A$,
\begin{equation}
\begin{aligned}
    \langle \tau_\xi \rangle := &\int_{\partial A } d\sigma_A (\mathbf{r})  \tau_\xi(\mathbf{r})  \frac{\hat{\mathbf{n}}_A(\mathbf{r}) \cdot(\mathbf{J}_{AB, \xi}-\mathbf{J}_{ss, \xi})}{\nu_{AB}(\xi)} \\
    =& {\langle q_- \rangle}_\xi /{\nu_{AB}(\xi)},\label{eq:MFPT}
\end{aligned}
\end{equation}
where $d\sigma_A(\mathbf{x})$ is the surface element on $\partial A$, $\hat{\mathbf{n}}_A$ is a unit normal vector pointing $\partial B$, and $w_\xi (\mathbf{r})=\hat{\mathbf{n}}_A(\mathbf{x}) \cdot(\mathbf{J}_{AB, \xi}-\mathbf{J}_{ss, \xi})/\nu_{AB}(\xi)$ is a normalized weighting factor that accounts for the proportion of the reactive flux relative to the steady state flux that lies along the normal direction on $\partial A$. The derivation is shown in Appendix \ref{AC}.  This expression reduces to the form derived in equilibrium where $q_+ = 1-q_- $ and $\mathbf{J}_{ss} = 0$.\cite{berezhkovskii2022relations} In one-dimensional barrier-crossing processes, we get the typical Hill relation\cite{hill2005free} or Kramers' theory\cite{hanggi1990reaction} where the reciprocal mean first passage time becomes a reactive flux over population of $A$ domain. However, for diffusive barrier crossings the relation between the mean first passage time and Kramers' flux is through the mean backward committor, $\langle q_-\rangle_\xi$ averaged over the steady state distribution $\rho$ with field $\xi$. This average includes the population of state $A$ but also includes a contribution from the intermediate domain, $\Omega_{AB}$, which can be large for diffusion-controlled reactions.


\begin{figure}
\centering
\includegraphics[width=9cm]{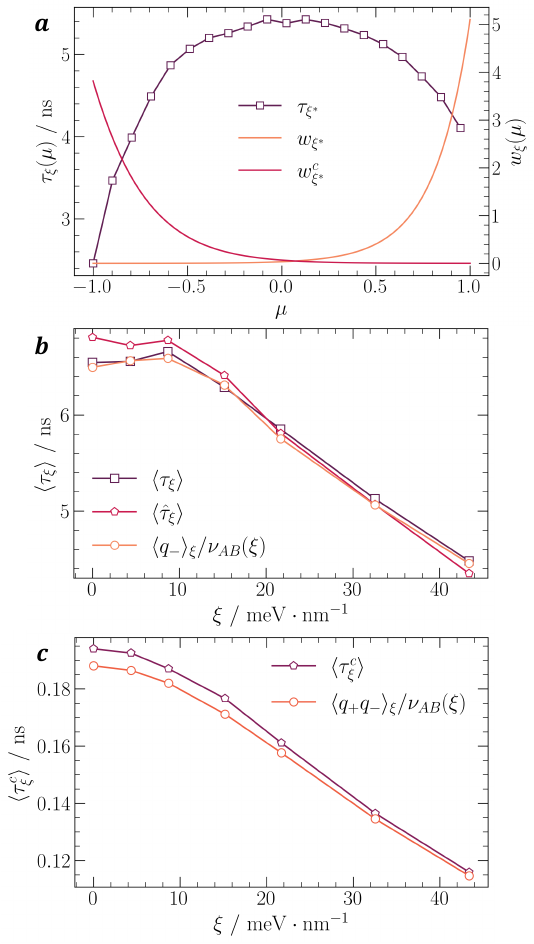}
\caption{\label{fig:3} a) Angle-dependent mean first passage time measured from simulations and two weighting factors $w_\xi$ and $w_\xi^c$ evaluated from analytic solutions in external force $\xi^*$. b) The averaged mean first passage times from analytic backward committor, molecular simulation of correlation the function $h_\xi'$, and mean backward committor over Kramers' reactive flux evaluated from analytic solutions. c) The averaged mean transition path time measured from simulations compared with the mean product of the forward and backward committors over Kramers' reactive flux evaluated from analytic solutions.}
\end{figure}

We measured the mean first passage time, $\tau_\xi(\mathbf{r})$ at each point in $\partial A$ from the molecular simulation and it is shown in Fig.~\ref{fig:3}a. Due to the spherical symmetry in our model system in equilibrium, $\tau_\xi(\mathbf{r})$ is uniform over the angle $\mu$. However, when the system is under a non-vanishing field, $\tau_\xi(\mathbf{r})$ is highest at the orthogonal direction to the field, and lowest at $\mu=-1$, the preferred side for recombination. The weighting factor $w_\xi(\mathbf{r})$ is computed from analytic solutions and it is also shown in Fig~\ref{fig:3}a. It is uniformly distributed in equilibrium, while out of equilibrium $\mu=1$ is strongly preferred where the AB-reactive flux has an opposite direction to the steady state flux.

The interpretation of the weighting factor in terms of the committor allows us to define a specific correlation function to measure the averaged mean first passage time directly from the simulation. We find,
\begin{align}
    w_\xi(\mathbf{r}) = -\hat{\mathbf{n}}_A(\mathbf{r}) \cdot D\rho\nabla q_- / \nu_{AB}(\xi)\, , \label{eq:ptweight}
\end{align}
where we used $q_- |_{ \partial A}=1$. If we consider another dividing surface $\partial A^+ =\{\mathbf{r} \ : \ |\mathbf{r}| =70 \ \mathbf{\AA} -\delta \} $, located slightly away from $\partial A$ by $\delta$, we can approximate the weighting factor as,
\begin{align}
    w_\xi(\mathbf{r}) \approx  \frac{D}{\delta \nu_{AB}(\xi) }\rho (1-q_-) \, ,\label{eq:ptcondition}
\end{align}
where it is proportional to the steady state probability conditioned on the trajectory that came from $\partial B$ not $\partial A$ from the past. Therefore, we can define a correlation function $h'_\xi(\tau)$, 
\begin{align}
    h'_\xi(\tau) = \frac{\langle I_{B}(\mathbf{r}_0 \in \partial A^+ )  I\left( \{ X_t\}  \right)  \rangle_\xi}{ \langle I_{B}(\mathbf{r}_0\in \partial A^+ )   \rangle _\xi } \, ,\label{eq:tcfconditioned}
\end{align}
where $I_B(\mathbf{r_0} \in \partial A^+)$ is an indicator function which is 1 if the trajectory starts from $\partial A^+$ and came from $\partial B$, not $\partial A$, in the past, and 0 otherwise. The averaged mean first passage time can be computed from 
\begin{align}
    \langle \hat{\tau}_\xi\rangle = \int_0^\infty dt \: {\color{black}\left [1-h_\xi'(t) \right ]}\, . \label{eq:meantau}
\end{align}
In Fig.~\ref{fig:3}b, we confirm that Eq.~\ref{eq:MFPT} is valid by comparing $\langle \tau_\xi \rangle$ which computed from $\tau_\xi(\mathbf{r})$ measured from simulations and the analytic weighting factor, with the analytic mean backward committor and $\nu_{AB}(\xi)$. Moreover, we confirm that the averaged mean first passage time directly measured from simulations with Eq.~\ref{eq:meantau} provides a consistent value. All measures of $\langle \tau_\xi \rangle$ confirm that the mean time to recombine decreases with $\xi$, in agreement with measures of the reactive flux in Fig.~\ref{fig:2}b. However $\xi$--dependent changes to the backwards commitor must be incorporated in order to find quantitative agreement between these measures of the characteristic time to react.  

Within a nonequilibrium steady state the mean transition path time, $\tau_\xi^c (\mathbf{r})$, has a connection to $\nu_{AB}(\xi)$ if it is properly weighted by accounting for persistent probability currents. We find that $\tau_\xi^c$ satisfies following differential equation,
\begin{align}
    \mathcal{L} (\tau_\xi^c q_+) = -q_+,  \label{eq:DEmtpt}
\end{align}
with boundary condition, $\tau_\xi^c |_{\partial B} = 0$. The proof is given in Appendix \ref{AD}. This expression is consistent with that reported in equilibrium when $\xi=0$.\cite{berezhkovskii2022relations} Following a similar route to derive Eq.~\ref{eq:MFPT}, we find the relationship between $\tau_\xi^c$ and $\nu_{AB}(\xi)$ in a nonequilibrium steady state,
\begin{equation}
\begin{aligned}
    \langle \tau_\xi^c \rangle := & \int_{\partial A } d\sigma_A(\mathbf{r}) \tau_\xi^c (\mathbf{r}) 
    \frac{\hat{\mathbf{n}}_A(\mathbf{r}) \cdot \left(\tilde{\mathbf{J}}_{ss, \xi} -\tilde{\mathbf{J}}_{AB, \xi}  \right)} {\nu_{AB}(\xi)}\\
    =& {\langle q_+ q_-\rangle}_\xi/{\nu_{AB}(\xi)},
    \label{eq:MTPT}
\end{aligned}
\end{equation}
where $\tilde{\mathbf{J}}_{AB}$ and $\tilde{\mathbf{J}}_{ss}$ are Kramers' reactive flux and steady state flux in a time-dual dynamics, respectively, $w^c_\xi=-\hat{\mathbf{n}}_A(\mathbf{r}) \cdot(\tilde{\mathbf{J}}_{AB, \xi}-\tilde{\mathbf{J}}_{ss, \xi})/\nu_{AB}(\xi)$ is a normalized weighting factor analogous to that appearing in Eq.~\ref{eq:MFPT} and it is shown in Fig.~\ref{fig:3}a. By definition of time-dual dynamics, $\tilde{\mathbf{J}}_{ss} = -\mathbf{J}_{ss}$, while, $\tilde{J}_{AB} \neq -J_{AB}$, which makes $w_\xi \neq w_\xi^c$. In equilibrium, the time-dual dynamics is identical to time-forward dynamics and $w_\xi = w_\xi^c$.\cite{berezhkovskii2022relations} 

The weighting factor for the mean first passage time, $w_\xi^c$ can be represented by committors,
\begin{align}
    w_\xi^c(\mathbf{r}) = \hat{\mathbf{n}}_A(\mathbf{r}) \cdot D\rho\nabla q_+ / \nu_{AB}(\xi)\, , \label{eq:ptweight2}
\end{align}
and it can be interpreted as hitting point distribution on the dividing surface $\partial A^+$ by the AB-reactive trajectories.\cite{vanden2006towards} Therefore, a collection of AB reactive trajectories sampled from the steady state molecular simulations is already weighted by $w_\xi^c$ naturally. The resulting averaged mean transition path time as a function of applied field is shown in Fig.~\ref{fig:3}c. We observe a decrease in $\tau_\xi^c$ with increasing field, indicating the field facilitate a trajectory passing through intermediate reactive domain if it is conditioned to be reactive without recrossing the boundaries.\cite{helfmann2020extending}

\section*{Conclusions}
In this work we have established generalizations of measures of the frequency of a reaction that evolves under diffusion control within a nonequilibrium steady state and probability fluxes. We have found that standard measures, like the Hill relation, require care in applying due to the existence of persistent probability currents. Further, measures afforded from simulation like correlation functions require strict definitions in order to be related to characteristic reaction times, due to the breakdown of a separation of timescales that accompanies diffusive dynamics. We applied this formalism within the context of ion pair recombination under applied electric field, and observed that all measures of the rate of the reaction increase with increasing field. We developed a thermodynamic interpretation of this flux increase, as well approximated by the work done on a free ion moving in the field over a typical transition path time. Future extensions of this formalism to molecular models of diffusion-controlled reactions under flow are ongoing. 

\section*{ACKOWLEDGMENTS}
 This work was  supported by NSF Grant CHE 2102314. S. M thanks Leonardo Coello Escalante and Jorge L. Rosa-Ra\'{\i}ces for extensive discussions. D.T.L. acknowledges support from an Alfred P. Sloan Research Fellowship.

\appendix
\renewcommand{\thefigure}{A\arabic{figure}}
\setcounter{figure}{0}  

\section{\label{AA}Details on Model Parameters}
We generated a reactive trajectory ensemble of ion--pair recombination by performing molecular dynamics simulations. A pair of singly-charged cation and anion is placed in the cubic box of length $L= 200\ \mathrm{\AA}$, interacting with each other by Coulomb interaction with dielectric constant of $\varepsilon=25$ and Lennard--Jones interaction dependent on the separation $r$ 
\begin{equation}
U_\mathrm{LJ}(r)=4 \epsilon \left [\left (\frac{\sigma}{r} \right)^{12}-\left (\frac{\sigma}{r} \right)^{6} \right ] \, ,
\end{equation}
with $\sigma = 2.0 \ \mathrm{\AA}$, $\epsilon = 0.01  \ \mathrm{kcal \cdot mol^{-1}}$ which is truncated and shifted at a cutoff of $100 \ \mathrm{\AA}$. Long--range electrostatics is included using the PPPM method.\cite{hockney2021computer} The equation of motion is propagated by the velocity Verlet algorithm with $10 \ \mathrm{fs}$ time step.\cite{limmer2024statistical} A Langevin thermostat is applied to fix the temperature at $300 \ \mathrm{K}$ with a damping constant of $150 \ \mathrm{fs}$, leading to the diffusion constant of $D_+ = D_- = 10 \ \mathrm{nm^2 \cdot ns^{-1}}$, which guarantees overdamped dynamics. $4 \mu s$-long 32 trajectories are sampled for each external field strength where initial $200 \ \mathrm{ns}$ are used for equilibration. The Debye--Huckel length of the system is $119.4 \ \mathrm{\AA}$, indicating the screening effect contributes minimally to the steady state distribution. All molecular dynamics simulations were performed with LAMMPS.\cite{thompson2022lammps}

\section{\label{AAA}General Solutions of Eq.~\ref{eq:sm}}
The general solutions of the Fokker--Planck equation in a steady state and the forward committor are provided from Refs.~\onlinecite{isoda1994effect, hong1978solution}. The angular function $T_j(\mu)$ is given as,
\begin{align}
    T_j(\mu) = \sum_{n=0}^\infty a_{jn}' \left( \frac{2n+1}{2} \right)^{1/2} P_n(\mu) \, ,
\end{align}
where $[(2n+1)/2]^{1/2}P_n(\mu)$ are orthonormalized Legendre polynomials. The coefficients $a_{jn}'$ are determined from the matrix $A$, 
\begin{align}
A_{mn} =
\begin{cases}
\frac{-m}{\sqrt{(2m-1)(2m+1)}}G & \text{if } n = m-1, \\
m(m+1) & \text{if } n=m, \\
\frac{-(m+1)}{\sqrt{(2m+1)(2m+3)}}G & \text{if } n = m+1, \\
0 & \text{otherwise},
\end{cases}
\end{align}
where $G=\beta \xi l_\mathrm{B}/2$ is unitless field strength and $a_{jn}'$ is the $n$'th element of the $j$'th eigenvector corresponding to eigenvalue of $\lambda_j$ ordered as $\lambda_0 < \lambda_1 <\lambda_2< \dots$. 

The radial function $Z_{2j}(r)$ is given as an infinite sum of modified Bessel functions of the first kind, denoted $J_{n+1/2} (1/\hat{r})$,
\begin{align}
    Z_{2j}({r}) =\left( \frac{2}{G} \right)^{1/2} \frac{\pi e^{-G\hat{r}/2}}{\sqrt{2}a_{j0}'} \sum_{n=0}^{\infty} (-1)^n a_{jn} J_{n+1/2} (1/\hat{r}) 
\end{align}
where $\hat{r}=2r/\ell_\mathrm{B}$ and $Z_{1l}({r})$ is related to $Z_{2l}({r})$,
\begin{align}
    Z_{1j}({r}) = Z_{2j} \left( \frac{2}{G{r}} \right) \, ,
\end{align}
by evaluating the former at a field weighted distance.

\section{\label{AB}Stochastic Work Bound}
\begin{figure}
\centering
\includegraphics[width=3.0in]{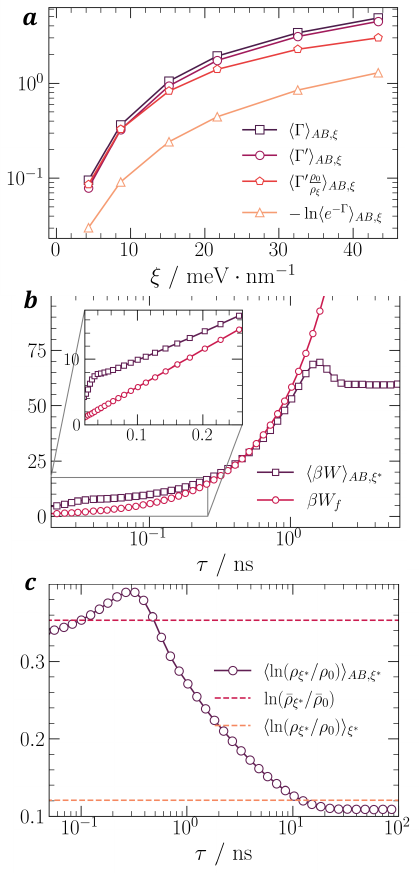}
\caption{\label{fig:s1}a) Measurement of the stochastic work bounds with the explicit correlation to the initial probabilities of AB reactive trajectories(pentagon) and by completely ignoring the initial probabilities(circle). Path averages are computed at $\tau^* = 2\tau^c_\xi$. b) Deviation of the conditioned work functional from the free work of non-interacting ion pairing system in $\xi^*$. Inset highlights the behavior in short times. c) Time dependence of the conditioned average of probability log-ratios $\xi^*$. The estimators at short-time and long-time limits are shown as dashed lines.}. 
\end{figure}

Jensen's inequality in Eq.~\ref{eq:jensen} provides a linear estimator readily accessible even from experimental observations of reactive trajectories. However, this approximation leads to a neglect of the correlation between the path variables $\Gamma'$ and the ratio of initial probabilities $\rho_0 / \rho_{\xi}$. To quantify the importance of this correlation, we applied Jensen's inequality partly to the path-dependent $\Gamma'$ term and compared it with the full exponential and full linear averages in Fig.~\ref{fig:s1}a. Moreover, we computed the bound by ignoring the initial probabilities contributions. The result shows the weighting factor $\rho_0/\rho_{\xi}$ reduces the bound by 60\% at $\xi^*$ while the effects are not significant in low field limit and when it is linearized. Therefore, the correlation from the initial points of trajectories becomes important in highly perturbed nonequilibrium diffusion-controlled processes, while it can be safely ignored in the linear response regime where the ratio of probabilities {\color{black}does} not deviate significantly from unity, or in high barrier crossing events where the memory of the initial location in reactive trajectory ensemble decays quickly.

The conditioned average work in Eq.~\ref{eq:Jensen} is compared with the free work of non-interacting ions in Fig.~\ref{fig:s1}b. The conditioned work shows a plateau  around the time scale of $1/\nu_{AB}(\xi)$ since trajectories spend comparable time in the ion--paired domain. The behavior of the conditioned work generally follows the free work until the plateau behavior, while some deviation occurs at short times. 
The time dependent log ratio of initial conditions are shown in Fig.~\ref{fig:s1}c. It is non-monotonic. At long times where the correlation of initial locations diminishes, the conditioned averages goes to the unconditioned averages, which we can directly compute from the analytic solutions of the steady state probabilities,
\begin{align}
    \lim_{\tau\rightarrow \infty} \left \langle \ln \frac{\rho_{\xi}}{\rho_0}\right \rangle_{AB, \xi} = 
   \left \langle \ln \frac{\rho_{\xi}}{\rho_0}\right \rangle_{\xi} \, ,
\end{align}
independent of the conditioning. At short times, the dominant contribution to the average comes from trajectories started from the vicinity of $\partial A$. Accordingly, 
\begin{align}
\lim_{\tau\rightarrow 0}\left \langle \ln\frac{\rho_\xi(\mathbf{r}_0)}{\rho_0(\mathbf{r}_0)} \right \rangle_{AB, \xi} \approx \ln \frac{\bar{\rho}_\xi}{\bar{\rho}_0} \, ,
\end{align}
where 
\begin{equation}
\bar{\rho}_\xi = \int_{\partial A}  d\sigma_A(\mathbf{r}) \rho_\xi(\mathbf{r}).
\end{equation}
This form of estimator is readily measurable from simulations by sampling reactive trajectories and taking a steady state average of an indicator function which is 1 if a trajectory is in the vicinity of $\partial A$ or 0 otherwise.

\section{\label{AC}Mean First Passage Time and Kramers' Flux}
We present the derivation of Eq.~\ref{eq:MFPT} for a general nonequilibrium steady state. For simplicity of notation, we introduce $\mathbf{b}(\mathbf{x}) = \beta D (\mathbf{F} +\mathbf{\xi} \hat{\mathbf{z}})$. By multiplying $q_- \rho$ on both sides of Eq.~\ref{eq:DEmfpt} and integrating over the complement of $B$ domain, denoted $B^C$,
\begin{align}
    \int_{B^C} d\mathbf{r} \: q_- \rho \mathcal{L} \tau_\xi   = -\int_{B^C} d\mathbf{r}  \rho q_- = -\langle q_-\rangle \, , \label{eq:integrated}
\end{align}
where we used $q_- |_{B}= 0$. Using the chain rule, and the definition of $\mathcal{L}$, the left hand side becomes,
\begin{equation}
\begin{aligned}
    q_- \rho & \mathcal{L} \tau_\xi  = q_- \rho (\mathbf{b} \cdot \nabla \tau_\xi + D\nabla^2 \tau_\xi ) \\
    & = q_- D\nabla \cdot (\rho  \nabla \tau_\xi) + q_-\mathbf{J}_{ss} \cdot \nabla \tau_\xi \, ,
\end{aligned}
\end{equation}
which follows from the definition of $\mathbf{J}_{ss} $.
To convert this into a divergence, we introduce an identity valid for arbitrary vector field, 
\begin{equation}
\begin{aligned}
    q_- D\nabla \cdot (\rho  \nabla \tau_\xi) &=  \tau_\xi \mathbf{J}_{ss} \cdot \nabla q_- \\
    &+ D\nabla \cdot ( q_- \rho \nabla \tau_\xi  - \tau_\xi \rho  \nabla q_-) \, , \label{eq:identity}
\end{aligned}
\end{equation}
which can be proved using the chain rule and the backward committor equation. Using the fact that the steady state flux is divergence--free, we get,
\begin{equation}
\begin{aligned}
    q_- \rho \mathcal{L} \tau_\xi &= D\nabla \cdot ( q_- \rho \nabla \tau_\xi  - \tau_\xi \rho \nabla q_-) + \mathbf{J}_{ss}   \cdot \nabla (q_-  \tau_\xi )  \\
     &= \nabla \cdot (D q_- \rho \nabla \tau_\xi  - D\tau_\xi \rho \nabla q_- + q_- \tau_\xi\mathbf{J}_{ss} ). \label{eq:divergence}
\end{aligned}
\end{equation}
The left hand side of Eq.~\ref{eq:integrated} becomes a form to which we can apply divergence theorem. The result of integration has two contributions, one is the surface integral on $\partial B$ and the other is the surface integral on $\partial A$ due to the discontinuity of $\nabla q_-$. The first one is zero since $\tau_\xi =0$ and $q_- = 0$ on $\partial B$. Therefore we get,
\begin{align}
    \int_{B^C} d\mathbf{r} \: q_- \rho \mathcal{L} \tau_\xi &= \int_{\partial A}d\sigma _A(\mathbf{r}) \, \hat{\mathbf{n}}_A \cdot D \tau_\xi \rho \nabla q_-    \\
    &=\int_{\partial A}d\sigma _A(\mathbf{r}) \tau_\xi \, \hat{\mathbf{n}}_A \cdot (\mathbf{J}_{AB} - \mathbf{J}_{ss} ) \, ,\nonumber
\end{align}
where we choose $\hat{\mathbf{n}}_A$ as a unit vector normal to the $\partial A$ directing $B$ domain and used the fact that $q_- =1$ on $\partial A$. Note that the right hand side is negative. Finally, dividing both sides of Eq.~ \ref{eq:integrated} with $\nu_{AB}$,  we get Eq.~\ref{eq:MFPT}.

\section{\label{AD}Equation for Mean Transition Path Time}
The mean transition path time $\tau_\xi^c(\mathbf{r})$ is the conditioned mean survival time of trajectory started from $\mathbf{r}$ escaping the intermediate domain $\Omega_{AB}$ through $\partial B$, not $\partial A$. The time, $\tau_\xi^c$ is defined as,
\begin{align}
    \tau_\xi^c(\mathbf{r})= -\int_0^\infty dt \ t \frac{d}{dt} \rho^c(\mathbf{r}, t)=\int_{0}^{\infty} dt \ \rho^c (\mathbf{r}, t)  \, , \label{eq:mtptdef}
\end{align}
where $\rho^c(\mathbf{r}, t)$ is the conditioned survival probability of trajectory started at $\mathbf{r}$. Specifically, given a set of trajectories started at $\mathbf{r}$ {\color{black}and committed to $\partial B$}, $\rho^c(\mathbf{r}, t)$ is the probability of observing a trajectory from the set that is not yet adsorbed to $\partial B$ at time $t$. In the second equality, we assume the boundary term vanishes, $\lim_{t \rightarrow \infty} t\rho^c(\mathbf{r}, t) = 0$ The conditioned probability $\rho^c(\mathbf{r}, t)$ is given as 
\begin{align}
    \rho^c(\mathbf{r}, t) =\int_{\Omega_{AB}} d\mathbf{r}' \rho(\mathbf{r}', t|\mathbf{r}) \frac{q_+ (\mathbf{r}')}{q_+(\mathbf{r})} \, ,
\end{align}
where $\rho(\mathbf{r}', t|\mathbf{r})$ is the time dependent solution of the Smoluchowski equation,
\begin{equation}
    \begin{aligned}
        \partial_t \ \rho(\mathbf{r}', t|\mathbf{r}) &= \mathcal{L}^{\dagger} 
        \rho(\mathbf{r}',t | \mathbf{r}),
    \end{aligned}
\end{equation}
with two absorbing boundary conditions on $\partial A$ and $\partial B$, $\rho(\mathbf{r}' , t|\mathbf{r}) |_{\mathbf{r'} \in A \cup B} =0$ and initial condition $ \rho(\mathbf{r}', 0|\mathbf{r}) = \delta(\mathbf{r}'-\mathbf{r})$. The Fokker--Planck operator, $ \mathcal{L}^{\dagger}$ acts on $\mathbf{r}'$.

We can multiply $q_+(\mathbf{r})$ to both sides of Eq.~\ref{eq:mtptdef} and apply $\mathcal{L}$, that acts on $\mathbf{r}$, on both sides,  
\begin{equation}
\begin{aligned}
    \mathcal{L}(q_+(\mathbf{r}) \tau_\xi^c(\mathbf{r})) &= \int_{\Omega_{AB}} d\mathbf{r}' \ \int_0^\infty dt \ \mathcal{L}  \rho(\mathbf{r}', t|\mathbf{r}) q_+(\mathbf{r}') \\
    &= \int_{\Omega_{AB}} d\mathbf{r}' \ q_+(\mathbf{r}')\ \int_0^\infty dt \frac{\partial}{\partial t}  \rho(\mathbf{r}', t|\mathbf{r}) \\
    &= -\int_{\Omega_{AB}}d\mathbf{r}' \ q_+(\mathbf{r}') \delta(\mathbf{r}'-\mathbf{r}) = -q_+(\mathbf{r}),
\end{aligned}
\end{equation}
we find Eq.~\ref{eq:DEmtpt}.

\section{\label{AE}Mean Transition Path Time and Kramers' Flux}
We present the derivation of Eq.~\ref{eq:MTPT} for a general nonequilibrium steady state. By multiplying $q_- \rho$ on both sides of Eq.~\ref{eq:DEmtpt}, we find,
\begin{align}
   q_- \rho \mathcal{L} \left( \tau^c_\xi q_+ \right)  = -q_- q_+  \rho. \label{eq:F1}
\end{align}
For the left hand side, we used Eq.~\ref{eq:divergence} for $\tau^c_\xi q_+$ to convert it into full divergence form, 
\begin{equation}
\begin{aligned}
    q_- \rho \mathcal{L} (\tau^c_\xi q_+) &= \nabla \cdot \{ D q_- \rho \nabla (\tau^c_\xi q_+) \\
    &- D(\tau^c_\xi q_+) \rho \nabla q_-
     + q_- (\tau^c_\xi q_+) \mathbf{J}_{ss} \} .
\end{aligned}
\end{equation}
Next, we integrate both sides over the $B^C$ domain. The right hand side Eq.~\ref{eq:F1} becomes a negative of the mean product of the forward and backward committors, $\langle q_+ q_-\rangle$ which is nonzero only in $\Omega_{AB}$. Using the divergence theorem, the left hand side of Eq.~\ref{eq:F1} becomes a surface integral. However, $q_- =0$, $\tau^c_\xi = 0$ on $\partial B$ so that the surface integral on $\partial B$ becomes zero. Similar to the mean first passage time, the only non-zero contribution comes from discontinuity of $\nabla q_-$ and $\nabla q_+$ on $\partial A$, which leads to,
\begin{equation}
    \begin{aligned}
        \int_{\partial A} d\sigma_A \ \tau^c_\xi D\rho  \ \mathbf{\hat{n}}_A \cdot \nabla q_+ = \langle q_+ q_-\rangle  \, , \label{eq:50}
    \end{aligned}
\end{equation}
where we used $q_-=1$, $q_+ =0$ on $\partial A$.

The left hand side of Eq.~\ref{eq:50} can be interpreted as the relative $AB$ reactive flux to the steady state flux in the time-dual dynamics. We note that the time-dual dynamics is an auxiliary dynamics with a different form of drift term $\mathbf{b}' = -\mathbf{b} + 2D\nabla \ln\rho$. The AB reactive flux in time-dual dynamics becomes,
\begin{equation}
\begin{aligned}
    \tilde{\mathbf{J}}_{AB} &= \mathbf{b}'\rho (1-q_+) - D\nabla  (\rho(1-q_+) ) \\
    &= (1-q_+) \tilde{\mathbf{J}}_{ss}   + D\rho\nabla q_+, 
\end{aligned}
\end{equation}
and Kramers' flux for time-dual dynamics normalizes the weighting factor since,
\begin{align}
    \tilde{\nu}_{AB}(\xi) &= \int_{\partial A} d\sigma_A(\mathbf{x}) \ \mathbf{\hat{n}}_A \cdot \tilde{\mathbf{J}}_{AB} \\
    &= \int_{\partial A} d\sigma_A \ D\rho \hat{\mathbf{n}}_A \cdot \nabla q_+ \, ,
\end{align}
where we used divergence-free property of $\mathbf{J}_{ss}$. As a final remark, the Kramers' flux for the time-forward and time-dual dynamics are identical\cite{vanden2006towards}, which proves Eq.~\ref{eq:MTPT}.

\section*{References}
\bibliographystyle{aipnum4-2}
\bibliography{main}

\end{document}